# Methane bursts as a trigger for intermittent lake-forming climates on post-Noachian Mars

Edwin S. Kite[1,*], Peter Gao[2,3], Colin Goldblatt[4], Michael A. Mischna[5], David P. Mayer[1,6], Yuk L. Yung[6].

1. Department of Geophysical Sciences, University of Chicago, Chicago, IL 60637.
2. NASA Ames Research Center, Mountain View, CA 94035.
3. Astronomy Department, University of California, Berkeley, CA 94720.
4. School of Earth and Ocean Sciences, University of Victoria, Victoria BC, V8P 5C2, Canada.
5. Jet Propulsion Laboratory, Pasadena, CA 91109.
6. U.S. Geological Survey, Astrogeology Science Center, Flagstaff, AZ.
7. Division of Geological and Planetary Sciences, California Institute of Technology, Pasadena, CA 91125. *e-mail: kite@uchicago.edu.

**Lakes existed on Mars later than 3.6 billion years ago, according to sedimentary evidence for deltaic deposition. The observed fluvio-lacustrine deposits suggest that individual lake-forming climates persisted for at least several thousand years (assuming dilute flow). But the lake watersheds' little-weathered soils indicate a largely dry climate history, with intermittent runoff events. Here we show that these observational constraints, while inconsistent with many previously-proposed triggers for lake-forming climates, are consistent with a methane burst scenario. In this scenario, chaotic transitions in mean obliquity drive latitudinal shifts in temperature and ice loading that destabilize methane clathrate. Using numerical simulations, we find that outgassed methane can build up to atmospheric levels sufficient for lake-forming climates, for past clathrate hydrate stability zone occupancy fractions >0.04. Such occupancy fractions are consistent with methane production by water-rock reactions due to hydrothermal circulation on early Mars. We further estimate that photochemical destruction of atmospheric methane curtails the duration of individual lake-forming climates to less than a million years, consistent with observations. We conclude that methane bursts represent a potential pathway for intermittent excursions to a warm, wet climate state on early Mars.**

Runoff on Mars after ~3.6 Ga was uncommon and episodic. Episodes of runoff are recorded by deltas and fans. Fan/delta watershed mineralogy shows limited aqueous weathering, and watershed topography lacks the slope-area scaling expected for prolonged fluvial erosion. Thus, mineralogy and geomorphology suggest runoff episodes were brief. Yet surprisingly, sediment and water mass balance calculations for ≲3.6 Ga precipitation-fed paleolakes do not suggest a palimpsest of catastrophic events. To the contrary, runoff production of 0.1-1 mm/hr and lake lifetimes of >3 Kyr (assuming dilute flow) requires sustained, non-catastrophic cycling of lake-water (e.g refs. 1-3) (Table S1). Catchments lack evidence for extensive leaching[4], and retain mafic minerals such as olivine, which dissolves quickly in water (Methods). Furthermore, late-stage fluvial incision into delta and fan deposits is uncommon. In summary, individual lake-forming climates lasted >3 Kyr but shut down rapidly.

**Drawing out the implications of intermittency data**

To draw out the implications of the intermittency data, we use a conceptual model of catchment response to a ~3 Ga wet episode (Fig. 1a). Consistent with models (e.g. Mischna et al. 2013), we assume that snow falls in low-latitude catchments when obliquity (φ) >40°. During a wet-climate anomaly, runoff from snowmelt transports sediment to build a fan/delta. This phase lasts >(3-10) Kyr (the product of delta volume and water:sediment ratio, divided by the energy-limited lake evaporation rate)[1-3] (Table S1). During this phase, erosion may expose mafic minerals (e.g. olivine) in sediment-source regions (Fig. 1a). As climate cools, meltwater production is insufficient to transport sediment, but still wets the active-layer soil and so dissolves olivine. The duration of this phase cannot exceed the olivine-dissolution lifetime, ~$10^6$ yr (refs. 5-6).

Hypotheses for the trigger of lake-forming climates should match constraints on the intermittency and cadence of those climates. Many existing hypotheses for the trigger of lake-forming climates underpredict lake lifetime (Fig. 1b). Individual volcanic eruptions only permit wet events of at most hundreds of years duration[7,8]. Models of ~3 Ga asteroid impacts predict <0.1 yr runoff[9]. Alternatively, a $H_2$-$CO_2$ greenhouse requires >0.15 Myr to remove $H_2$ at the diffusion-limited rate[10]; this is marginally consistent with data, but requires a brief >$10^7$ $km^3$ pulse of late-stage volcanism (or clathrate release; ref. 11) to provide $H_2$. Recently, limit cycles involving rapid deglaciation and rapid carbonate formation have been proposed to explain >3.6 Ga lake-forming climates[12]. Such a limit cycle is implausible for ≲3.6 Ga lakes because post-Noachian soil thicknesses and erosion rates provide insufficient cations for rapid weathering drawdown of the atmosphere[13], and because the hypothesized carbonates would reside near the modern surface in conflict with spectroscopic constraints[14].

**Methane bursts as a trigger for intermittent lake-forming climates**

An alternative trigger for lake-forming climates is chaotic transitions in Mars' mean obliquity. These transitions are large (10-20°), brief (often ≲$10^7$ yr), and infrequent: transported to a random point in Mars' history, one would expect to find oneself in a 0.5 Gyr-long interval of continuously high (or low) Myr-mean **φ** (Fig. 2). The brevity and large time interval of mean-**φ** transitions matches the brevity and rarity of lake-forming climates. Moreover, mean-**φ** transitions cause latitudinal shifts in temperature, which destabilizes ice/snow (e.g., ref. 15). Thus **φ** shifts can increase the amount of water in the atmosphere, favor cirrus-cloud warming[16], and prime surface snowpack for runoff[15].

During a large shift in mean **φ**, the subsurface will undergo correspondingly large changes in pressure (as surface ice and ground ice migrate) and temperature. At some latitudes this will destabilize $CH_4$-clathrate, yielding $CH_4$ gas[17]. $CH_4$-clathrate breakdown involves a >14% reduction in solid volume, and we assume fractures allow methane gas released at ≤100m depth to reach the surface in $\ll 10^4$ yr (ref. 18). In our scenario, the ultimate source of $CH_4$ is hydrothermal circulation (e.g. serpentinization) early in Mars history (e.g. ref. 19). The $CH_4$-production stoichiometric upper limit (for hydrothermal reactions) is $>10^4$ × greater than the amount needed to shift planetary climate (Methods). Methane-saturated fluids will deposit clathrate on approach to a cold surface. As Mars cools, the hydrate stability zone (HSZ) expands. Methane will diffuse out of the HSZ only slowly, but once destabilized, $CH_4$-clathrate dissociates geologically quickly[20].

Mean-**φ** transitions can lead to build-up of millibars of methane in Mars' atmosphere. To show this, we used calculations of Mars' spin and orbit [21], output from a Global Climate Model[15], a parameterization of the greenhouse effect of $CH_4$ (ref. 22), and a photochemical model of $CH_4$ destruction[23], in order to drive a model of $CH_4$-clathrate stability in Mars' subsurface (Methods). We used mobile ice+dust overburden of ~40 m thickness[24]. The fraction *f* of the HSZ that is occupied by hydrate is a free parameter. Example output is shown in Fig. 3. Following model spinup, little happens for ~0.2 Gyr. $CH_4$ released to the atmosphere during quasi-periodic orbital change[17] is maintained below radiatively significant levels by UV photolysis[25]. Then, a mean-**φ** shift occurs, swiftly destabilizing $CH_4$-clathrate (Fig. 3). $CH_4$ release temporarily overwhelms photolysis by UV (we use a ~3.0 Ga UV flux; ref. 26); $CH_4$ accumulates in the atmosphere. Our photochemical modeling (Methods) shows that the Mars atmosphere $CH_4$-enrichment episode duration is set by UV photon supply, the $CH_4$ /$CO_2$ ratio, and other photochemical effects, and is $10^5$-$10^6$ years. This duration allows for multiple orbitally-paced pulses of runoff in a given lake basin, consistent with data[3], and satisfies the lake duration constraint. $CH_4$ peaks early in the episode and declines gradually. Because the $CH_4$-clathrate reservoir is recharged slowly if at all, the $CH_4$-burst mechanism also satisfies the olivine-dissolution constraint.

$\gtrsim$1% methane can switch the Mars system from zero meltwater production to a lake-forming climate. 1% of $CH_4$ added to a ~1 bar $CO_2$ atmosphere in a clear-sky radiative-convective calculation boosts temperature by 6K (ref. 22). These corresponding $CO_2$ pressures are consistent with proxy data[27-28], assuming lakes were ice-covered. The boosted temperatures are high enough for perennial ice-covered lakes to form[29-30]. In our calculations $CH_4/CO_2 \leq 0.1$, so photochemical production of $C_2H_6$ is minor and antigreenhouse haze cannot form. However, abiotic hydrothermal reactions produce $C_2H_6$ with $10^{-3}$-$10^{-1}$ the efficiency of $CH_4$ (ref. 31), and $C_2H_6$ partitions readily into clathrate[18]. The greenhouse effect of even 1% $C_2H_6/CH_4$ would be radiatively significant[32-33]. Although $CH_4$ photolysis yields $H_2$, the additional warming is modest. The likely presence of clouds would moderate total warming, but only by 14-30% (ref. 34).

**Methane bursts link subsurface and surface hydrology**

$CH_4$-induced surface warming swiftly destabilizes $CH_4$-clathrate at greater depth, which releases additional $CH_4$. This $CH_4$-release feedback greatly increases total $CH_4$ warming for $f > 0.04$; for $f < 0.04$, it is difficult to trigger a lake-forming episode. In addition, during a lake-forming event, lake-bottom temperature rises to >273K even for ice-covered lakes. This warming destabilizes sub-lake clathrates. Subsequent $CH_4$ degassing (e.g. via mud volcanoes) adds to atmospheric $CH_4$. The largest proposed paleolake on Mars is $10^6$ km$^2$ (ref. 35) and seas as large as $2.3\times10^7$ km$^2$ have been suggested (e.g. ref. 36). Because the lake-bottom warming is long-lived, sub-lake pore ice melts to open permeable conduits (through-taliks) to the deep hydrosphere.

There are other mechanisms by which mean-$\varphi$ transitions could drive lake formation by linking surface and subsurface hydrology[37]. For example, ice unloading could promote hydrofracture discharge of overpressured aquifers. Clathrate decomposition, e.g. driven by $\boldsymbol{\varphi}$ changes, might directly trigger outflow channels[37], and chaos terrain formation could also release $CH_4$ : individual chaos-terrains have volumes up to $10^5$ km$^3$. Chaos terrain formation could be associated with shallow magmatic intrusions, which might themselves destabilize $CH_4$ .

**Lake-forming climates in the context of Mars history**

Because widely-spaced $CH_4$ bursts are possible, we hypothesize that the ~3 Ga lake-forming climate may be a late echo of the more-intense ~4 Ga climate upswing that cut valley networks and filled inland seas[38]. For example, atmospheric collapse could drive ice sheets from highlands to poles[39], depressurizing sub-ice clathrate. Conversely, >5 widely-separated lake episodes would be inexplicable by $CH_4$ bursts alone.

In our model, surface climate $\lesssim$3.6 Gyr ago is driven by $CH_4$ produced during earlier water-rock reactions[11]. >3.6 Ga serpentinization has been documented from orbit[40], and data suggest cool mid-crustal temperatures that are consistent with pervasive hydrothermal circulation[41-42]. The abundant $CH_4$ predicted by our model could be tested with better constraints on permeability, alteration extent, and fluid chemistry in ancient deep aquifers[43]. Furthermore, the $CH_4$-clathrate reservoir on Mars should never completely vanish. Present-day $CH_4$ outgassing is predicted[44]. $CH_4$ outgassing has been reported from ground-based and rover instruments (e.g. ref. 45). Our model indicates that (after many chaotic-obliquity shifts) ancient clathrate should be closest to the surface in dusty longitudes at 30-50°N. The ExoMars orbiter may decisively test modern outgassing[46].

Our model of uncommonly wet climates on post-3.6 Ga Mars does not account for the lesser amounts of liquid water needed to explain the prolonged accumulation of sedimentary rocks near Mars' equator (e.g. 47-48). While $CH_4$ bursts can explain the cadence of lake-forming climates, the problem of accounting for these sedimentary rocks remains open[39,49].

**References.**

**Acknowledgements.** We are grateful for input from D.E. Archer, J.C. Armstrong, B.L. Ehlmann, V.E. Hamilton, A.D. Howard, R.P. Irwin III, M.C. Palucis, D. Stolper, R.M.E. Williams, and R. Wordsworth. We thank J.F. Kasting and A.G. Fairén for useful reviews. Part of the research was carried out at the Jet Propulsion Laboratory, California Institute of Technology, under a contract with the National Aeronautics and Space Administration. We acknowledge the University of Chicago's Research Computing Center and financial support from NASA (NNX16AG55G, NNX15AM49G).

**Author contributions.** E.S.K. designed research; M.M., Y.Y., and D.P.M. contributed new models, model output, and analyses; E.S.K., C.Z.G., and P.G. carried out research; and E.S.K. wrote the paper.

**Additional information.** Correspondence and requests for materials should be addressed to E.S.K. (kite@uchicago.edu)

**Competing financial interests.** The authors declare that they have no competing financial interests.

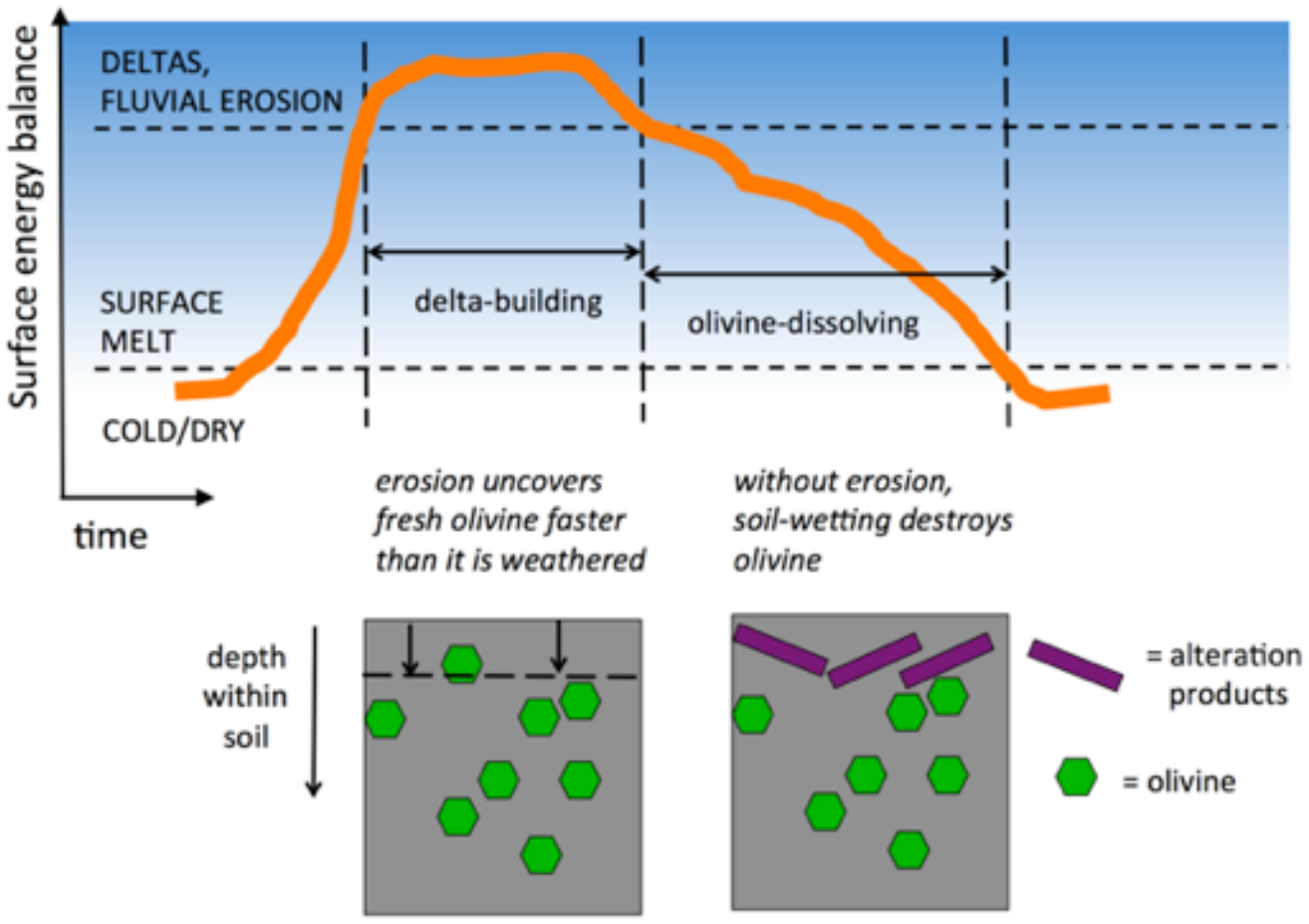

**(b)**

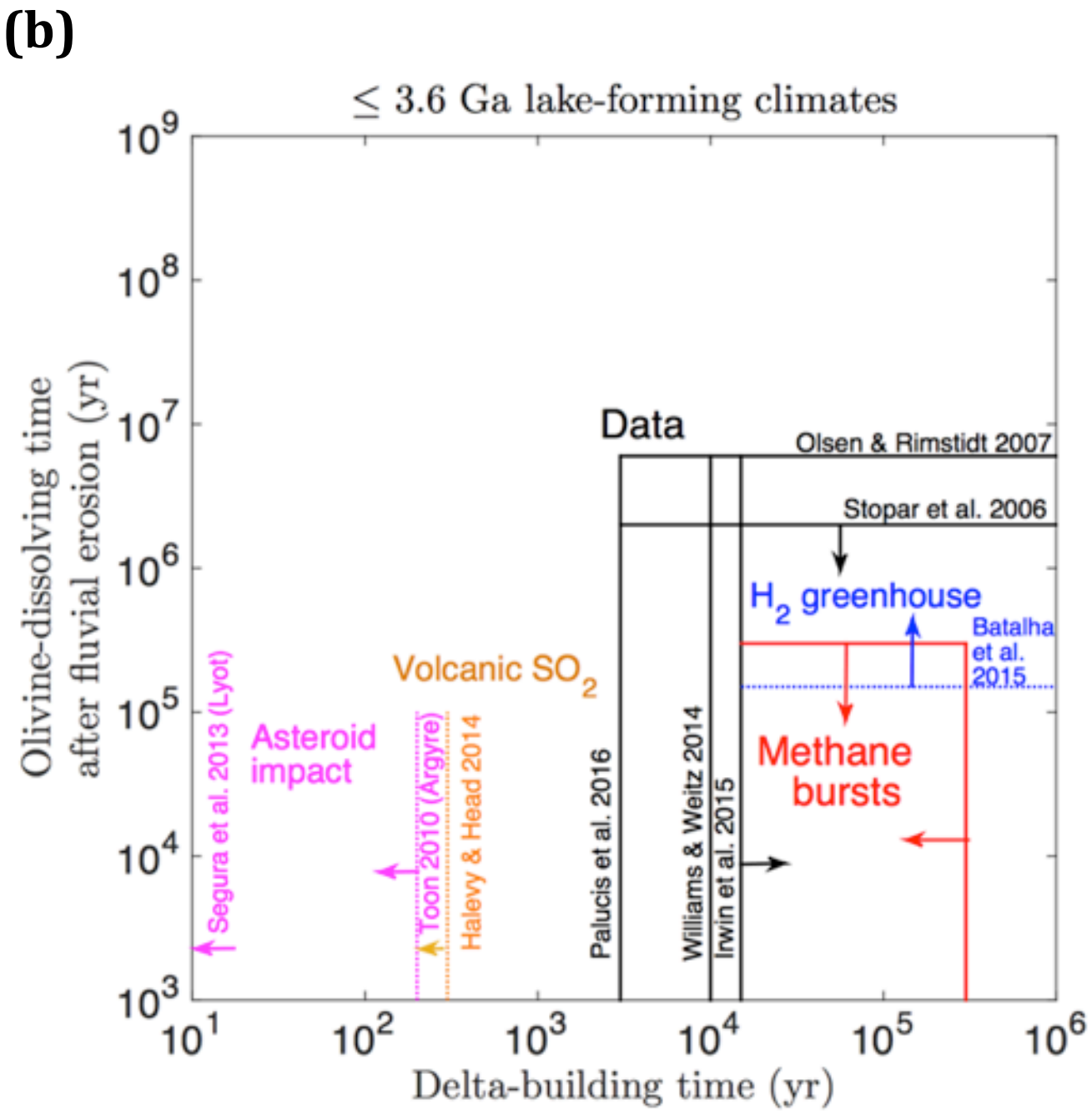

**Fig. 1.** Geologic constraints on the duration and shutdown-time of lake-forming climates. (a) Schematic showing how the delta-building timescale and the olivine-preservation timescale constrain duration and shutdown-time for a lake-forming climate episode. Olivine constrains soil-wetting after fluvial erosion ceases. (b) Geologic constraints (black lines) compared to models for the trigger mechanism of lake-forming climates (colored lines).

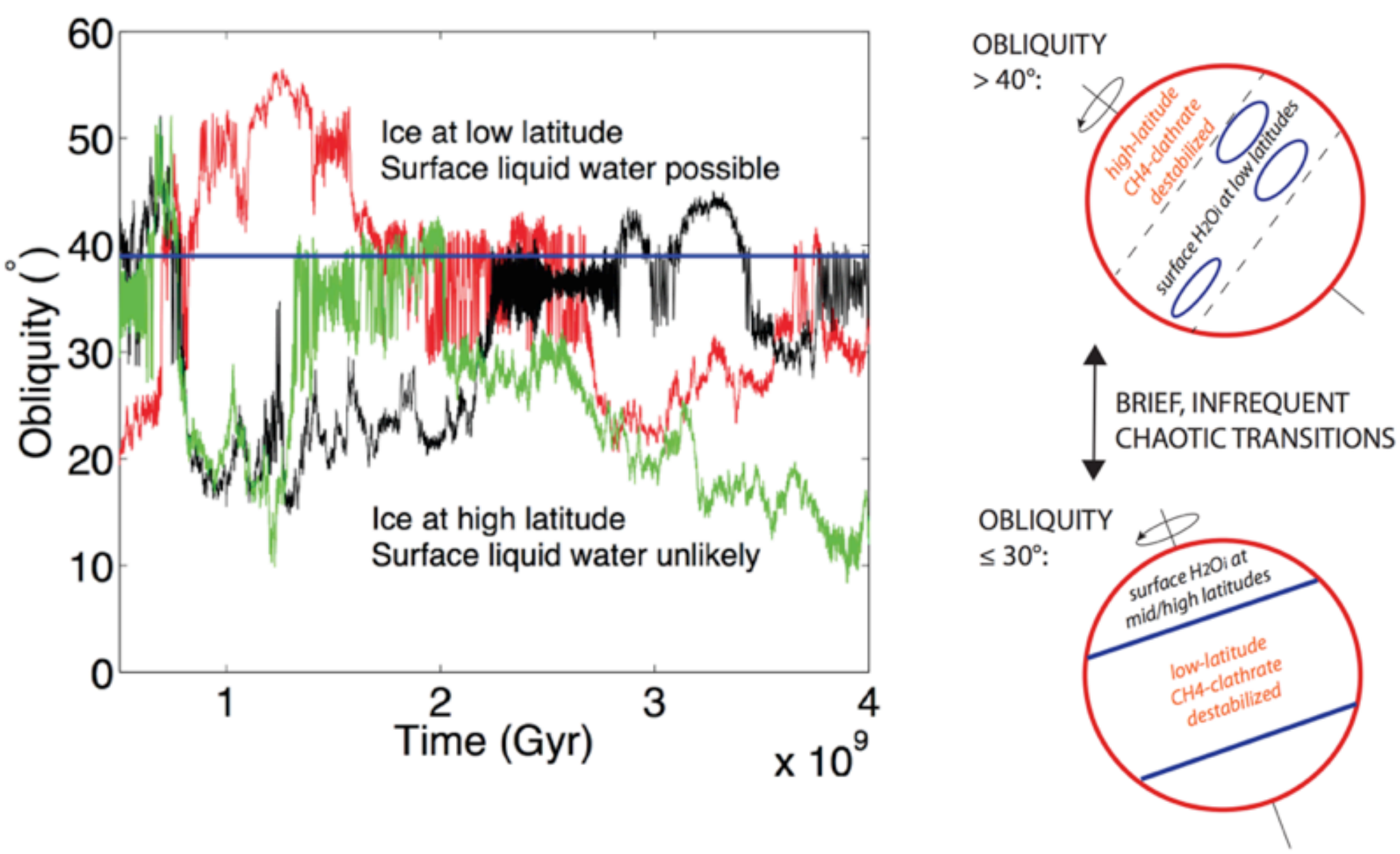

**Fig. 2.** Range of obliquity trajectories possible for Mars, and their likely climate effects. *Left:* Examples of possible, equally-likely, orbital histories for Mars. *Right:* Schematic showing effect of obliquity change on surface-ice distribution.

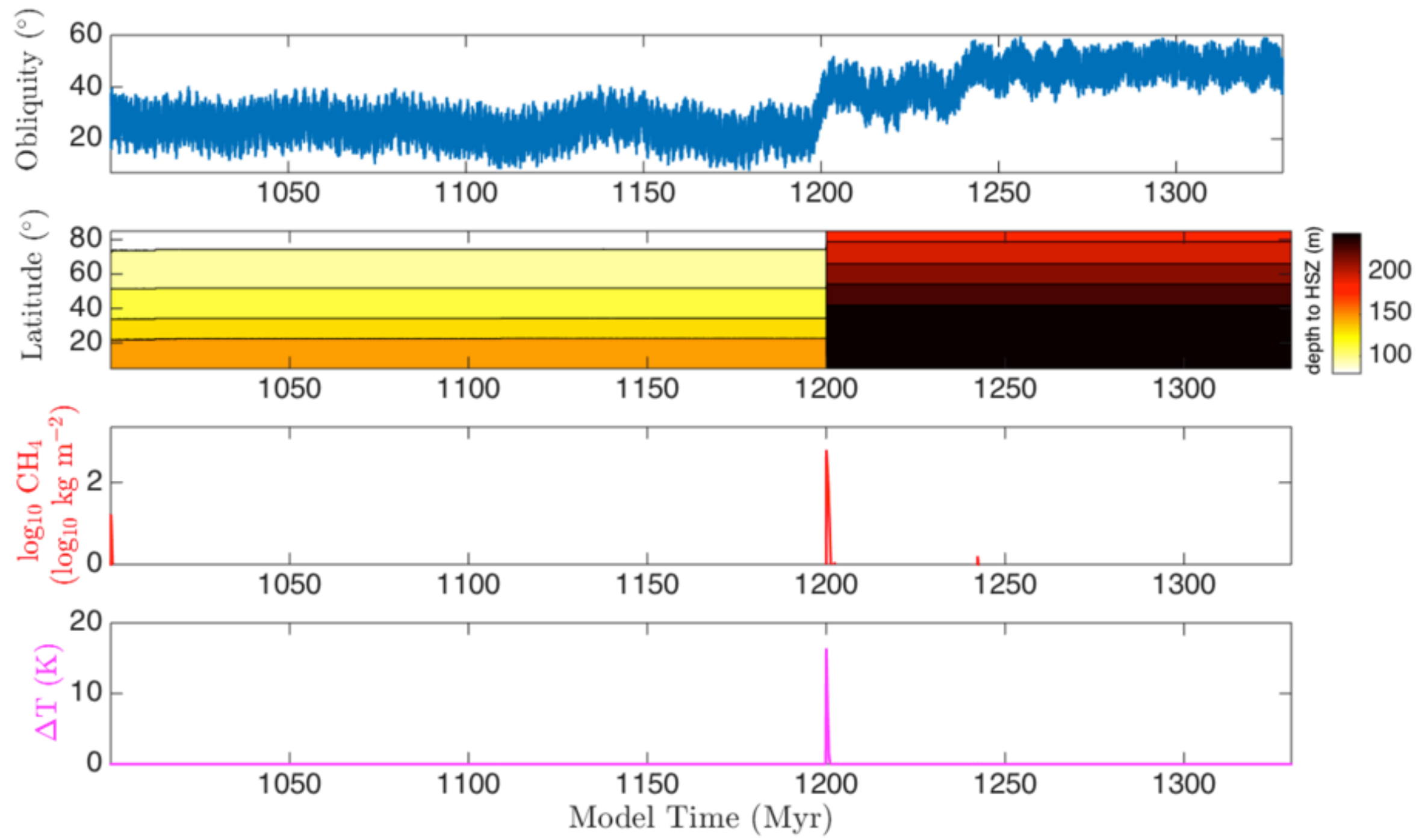

**Fig. 3.** Triggering of a $CH_4$-enabled lake-forming climate on Mars. Model time is arbitrary. (a) Example obliquity forcing (preceded by 0.3 Gyr of continuously

low obliquity). (b) Depth to the top of the clathrate-hydrate stability zone. Darkening denotes clathrate destabilization. (c) Atmospheric $CH_4$ column mass ($f$ = 0.045). (d) Temperature change (from parameterization of Wordsworth et al. 2017). Temperature change is sufficient to produce an Antarctic-Dry-Valleys like climate. Lake-forming event lasts ~300 Kyr (Fig. S5a).

**Methods.**

**Geologic Synthesis.**

For many low-latitude fluviolacustrine deposits on Mars, geologic data indicate mid-Hesperian to Early Amazonian age[1-3,50-57]. This corresponds[58] to ~3 Ga (or ~2 Ga using an alternative chronology[59]). These deposits postdate the ~3.8 Ga highland valley networks of Mars[50], which are also hydrologically distinct from the younger fluviolacustrine deposits[60]. Published data are consistent with the globally-distributed deposits having been caused by 1-2 intervals of delta-building. The water source for the features considered in this paper was precipitation (rain or snowmelt). Precipitation is indicated by spatially clustered watersheds with channels extending near ridgelines[61]. Snowmelt is a reasonable hypothesis for ~3 Ga deposits, although the reasoning set out in this paper does not rule out rainfall. Other ~3 Ga paleochannels and fluvial deposits (not considered here) are more ambiguous, and might be formed either by precipitation runoff[62-63] or by localized water sources[64-66].

The delta-forming duration $\tau_1$ (duration of fluvial sediment transport) must exceed

$$\tau_1 > V_d(V_w/V_s)/E\,A_p \qquad (1)$$

where $V_d$ ($m^3$) is measured delta volume, $V_w/V_s$ is water:sediment ratio, $E$(m $yr^{-1}$) is the sum of evaporation rate (constrained by energy balance) and infiltration rate, and $A_p$ ($m^2$) is lake area[1-2] (Table S1). Energy balance limits evaporation rate to <1 m $yr^{-1}$, and long-term average infiltration rate is likely small[1]. Deposit morphology suggests dilute flows (debris-flow deposits are less common[67]). Most authors therefore assume a dilute $V_w/V_s$ ratio ≥ $10^3$, similar to Earth data[68]. Long minimum lake lifetimes inferred from delta volumes (Table S1) are consistent with minimum runoff durations calculated from the energy-balance limit on snowmelt runoff production[49] combined with alluvial-fan volumes[67]. Because water demands (column meters) of both deltas and alluvial-fans exceed the plausible thickness of preexisting snowpack, precipitation recharge of water source areas must have occurred during the wet event (i.e., a hydrologic cycle)[1,67].

Assuming runoff from snowmelt, runoff rate is directly related to surface energy balance. The surface energy balance difference $J$ ($W/m^2$) between the energetic threshold for soil wetting (no runoff), and the same threshold for fluvial sediment transport, is

$$J = \rho L(Q/A + I + E_e) \qquad (2)$$

where $\rho$ is liquid water density (1000 kg m$^{-3}$), $L$ is latent heat of melting snow/ice (334 kJ kg$^{-1}$), $Q$ is river paleodischarge (m$^3$ s$^{-1}$), $A$ is drainage area in m$^2$ ($Q/A$ is "runoff production"), $I$ is infiltration rate (mm hr$^{-1}$), and $E_e$ is excess evaporation (mm hr$^{-1}$). A lower bound on $J$ is obtained by setting $I$ and $E_e$ to zero; then $J = \rho LQ/A$. For ≲3.6 Ga precipitation-fed channels, $Q/A$ is estimated as ~0.1–0.2 mm hr$^{-1}$ for Saheki[69], 0.03 – 0.4 mm hr$^{-1}$ for Peace Vallis[70], and 0.1-0.3 mm hr$^{-1}$ for Eberswalde[1]. Taking 0.2 mm hr$^{-1}$ as representative, $J$ = 20 W m$^{-2}$. These modest runoff requirements are consistent with an Antarctic-Dry-Valleys-like climate. An Antarctic-Dry-Valleys-like climate permits large lakes[29-30].

We use olivine persistence as a constraint on the duration of soil-wetting climates (Fig. 1). Olivine is present in many Mars delta and alluvial-fan watersheds. Specifically, (1) Observatoire pour la Minéralogie, l'Eau, les Glaces et l'Activité (OMEGA) shows olivine[74] in some alluvial fan watersheds. (2) Thermal Emission Spectrometer (TES) data shows widespread olivine[75] including in alluvial fan watersheds. (3) Thermal Emission Imaging System (THEMIS) decorrelation stretches indicate olivine[76] in alluvial fan watersheds. (4) The Compact Reconnaissance Imaging Spectrometer for Mars (CRISM) OLINDEX3 parameter, which was designed as an indicator of olivine[77], shows high values in many alluvial fan watersheds. CRISM olivine detections in alluvial-fan watersheds include Robert Sharp crater[78] and Saheki (Fig. S1). Olivine persistence sets an upper limit on the duration of soil wetting by olivine-dissolving fluids. Olivine-dissolution data indicate olivine lifetimes <(2-6) Myr for $T$~278 K, for a pH corresponding to pure water pH equilibrated with 60 mbar $CO_2$, and including a 100× lab-to-field correction[5-6]. Thicker atmospheres, as required for most warming mechanisms, give lower pH and thus shorter lifetimes. However, buffering by rock dissolution increases pH and thus olivine lifetimes[79]. Some calculations[80] give olivine lifetimes in fluids at Mars' surface as short as 10 yr. Short olivine-inferred water durations are consistent with short water durations inferred from (1) persistence of hydrated amorphous silica[81], (2) commingling of unaltered olivine with sulfates in the Peace-class rocks at Gusev[82], (3) persistence of jarosite[83], and (4) near-isochemical alteration of Bradbury group materials at Gale crater[84], among other methods. We assume that infiltrated water is present in soil throughout a wet season. This is reasonable because runoff generation from snowpack is extremely difficult unless snowpack reaches thermal maturity. Thermal maturity requires that average temperature during the warm month is near freezing[85]. Additionally, infiltration, and latent heat release, protect water from complete freezing.

Although we focus on ≲3.6 Ga lake-forming climate(s) in this study, the ~3.8 Ga lake-forming climate is also characterized by a relatively short-lived interval of intense fluvial sediment transport[38,71-72] (see also ref. 73).

Our model includes a $H_2O$ ice overburden that shifts with obliquity. Snow accumulation at latitude <45° at φ>40° is supported by all Mars climate models[15,39,49,86].

Snow accumulation at latitude <45° at $\varphi$>40° is also supported by observations of equatorial relict ice and glacial moraines[86]. This latitudinal shift in snow distribution is caused by the increase in polar summer insolation at high obliquity. At high atmospheric pressure, snow/ice may be present at low latitudes regardless of $\varphi$, but water ice stability patterns still show latitudinal shifts[39] with $\varphi$. Laterally-extensive midlatitude volatile-rich layers that migrate under $\varphi$ control were ~32 m thick based on relict Amazonian deposits[87], or (44±23)m thick based on pedestal craters[24].

**Assessment of previously-proposed trigger mechanisms.**

- Volcanic $SO_2$: The $SO_2$-greenhouse model of ref. 7 predicts wet events of duration ~30 yr. This hypothesis struggles to match minimum lake-lifetime constraints, and $SO_2$-outgassing may in fact induce net cooling[8].
- Climate change triggered by impact energy: Impact-triggered models for post-3.6 Ga wet climates must satisfy the geologic constraint of modest precipitation-sourced erosional modification of the six largest post-3.6 Ga craters on Mars[88]. Ref. 16 proposes a metastable impact-triggered wet climate sustained by cloud forcing. Such a climate can sustain temperatures above 273K on annual average, but only with unrealistic total cloud cover[89]. However, the model could generate seasonal melting with more realistic cloud-cover assumptions[89]. Ref. 9 states that a metastable warm/wet climate can be attained from the impact of an 8 km-radius asteroid. Their maps do not show rain at the impact location itself, which is intriguingly consistent with post-3.6 Ga Mars data[88].
- Impact delivery of volatiles: Comets have ~1% $CH_4$ (ref 90.) One of the largest impact craters on Mars with age <3.6 Ga is Lyot (ref. 88). Supposing Lyot to have been formed by a comet, the impact would have delivered <0.01 mbar $CH_4$ (radiatively negligible).

**Obliquity Simulations.**

Mars obliquity ($\varphi$) is quasi-periodic on $<10^6$ yr timescales but chaotic on $\gtrsim 10^8$ yr timescales, ranging from 0-70° (ref. 91). These large changes have correspondingly large effects on climate[15]. To generate realistic possible $\varphi$ histories for ~3 Ga Mars, we first generated an ensemble of solar system simulations using the `mercury6` N-body code[92]. We added $\varphi$/precession tracks in post-processing using (ref. 93). We generated randomness by shifting Mars's initial position, and by randomly selecting initial obliquities from the probability distribution functions of ref. 91. After the tracks have diverged from their initial conditions, each track (and each time interval of a given track) is an equally good estimate of ~3 Ga Mars behavior. Our obliquity runs show chaotic transitions[93] in mean $\varphi$. Transitions are separated by long periods during which mean $\varphi$ does not vary greatly, consistent with previous work[94-95]. Our eccentricity pdf agrees with that of ref. 91. Our φ pdf is unimodal, peaking at ~40°, and with a shape close to that of ref. 91.

**Surface Temperature Modeling.**

We calculate surface temperature as a function of obliquity, latitude, $CO_2$ partial pressure, and $CH_4$ partial pressure. Our starting point is a grid of output from 20 runs of a $CO_2$ only GCM (derived from ref. 15), which were carried out assuming a solar luminosity 75% that of today's Sun, $pCO_2$ = {6, 60, 600, 1200} mbar, and **φ** = {15°, 25°, 35°, 45°, 60°}. We zonally-average and time-average the surface temperatures. We adjust results upwards by a fixed amount to match the results[96] of the LMD GCM, because the LMD GCM includes cloud and $H_2O$(v) effects that are absent in our GCMs. The LMD GCM predicts equatorial, datum-elevation, $CH_4$-free temperatures of ~245K. For a given $pCO_2$, we interpolate in the grid of adjusted GCM results using our obliquity tracks to interpolate the surface temperature as a function of time and latitude. (The effects of varying eccentricity and longitude of perihelion are neglected, which has the indirect effect of stabilizing clathrate at low latitude.) These $CH_4$-absent temperatures drive initial $CH_4$ destabilization. After non-negligible $CH_4$ has entered the atmosphere, we add a uniform temperature offset to take account of $CH_4$-$CO_2$ collision-induced absorption[22]. This calculation includes the surface-cooling effect associated with the absorption of near-infrared sunlight by methane[22]. Our simple approach to calculating greenhouse forcing is appropriate for this study, because our goal is to explain runoff intermittency (not absolute temperature, not latitudinal gradients, not the existence of runoff).

**Clathrate Modeling.**

*Charge-up.* Early Mars had active magmatism, initially high geothermal heat flow, likely a large water inventory[97], and a basaltic/ultramafic crust. Therefore, the amount of $CH_4$ produced by serpentinization early in Mars history is potentially very large[98-99]. Whether this $CH_4$-production potential was realized depends on details of catalyst distribution and crustal permeability, which are poorly known even for Earth. Our $CH_4$-burst scenario requires no more than 0.0001× of the stochiometric $CH_4$-production upper limit. Mars meteorites have ~(15-20) wt% $FeO_T$, i.e. 10-13 wt% Fe. Most of the Fe in the upper crust has FeO oxidation state. Assuming 1 electron per oxidized Fe, the $CH_4$-production upper limit is 7% of the mass of the Fe. Multiplying by an assumed mid-crustal alteration zone thickness of 5 km and density 3000 kg/m$^3$ yields an electron-based stoichiometric upper limit of ~$10^2$ bars. In practice, the limit to $CH_4$-production would more likely be set by C availability. Mars crust production was extended over a long period including times during which the surface would have been cold. Therefore, $CH_4$ produced by water-rock reactions (for example by serpentinization and Fischer-Tropsch Type reactions; refs. 30,100) would have been trapped on approach to the cooling surface (e.g. beneath ice sheets or primordial seas) as clathrate (Fig. S3). $CH_4$ accumulates in ~$10^7$ yr by cycling of $CH_4$-saturated water through the Hydrate Stability Zone (HSZ), or more quickly by bubble exsolution[101]. The fraction of HSZ volume that is occupied by clathrate ($f$ in our model) must be divided by porosity to obtain the fraction of pore space that is occupied by clathrate. We assume a porosity of 0.3 (Lunar porosity is ~0.25)[102]. The extent to which pore space is filled on Mars by abiotic methane clathrate is unknown[103]; on Earth biogenic methane clathrate fills ~3% of available pore space[104].

*Release.* $CH_4$ trapped in clathrate is retained for up to Gyr. Ref. 105 cites theoretical calculations[106] for which the diffusivity of $CH_4$ in the clathrate lattice ($D_{CH4}$) is given by

$$D_{CH4}(T) = 0.0028 \times X_{CH4} \exp(-6.042 \times 10^{-13} / kT) \text{ cm}^2 \text{ s}^{-1} \quad (3)$$

where $k$ is Boltzmann's constant. Setting $X_{CH4}$ = 0.03 (where $X_{CH4}$ is the fraction of unoccupied clathrate-lattice cages) and $T$ = 270K (worst case for $CH_4$ loss) yields $8 \times 10^{-16}$ m$^2$ s$^{-1}$. In this case, 3 Gyr will allow approximately 10m of clathrate to be de-methanated by $CH_4$ loss, which is not important for our purposes. Plausible increases[20] to $10^{-14}$ m$^2$ s$^{-1}$ do not alter this qualitative conclusion. $CH_4$ clathrate that is moved out of the P-T range of $CH_4$-clathrate stability will outgas $CH_4$ geologically quickly[107].

We calculate $CH_4$-clathrate stability assuming that thermal equilibrium is reached at each 1 Kyr timestep. This is reasonable because the depth of clathrate destabilization in our model is ≲250m. Tests using a 1D scheme (tracking temperature as a function of depth) showed no qualitative difference in behavior. The latent heat of clathrate dissociation is ignored; this is acceptable because the thermal forcing of interest (from orbital variations) varies slowly compared to the speed of lowering of the clathrate table with or without latent-heat buffering. Geothermal heat flux is 0.03 W m$^{-2}$. The model is spun up with zero clathrate release for 5 Myr.

Regolith density (on top of the HSZ) is 2000 kg m$^{-3}$. A surface layer of 44m of ice (density 910 kg m$^{-3}$) is assumed poleward of 30° for $\varphi < 40°$ (ref. 24). When $\varphi > 40°$, this ice sublimates at 0.3 cm yr$^{-1}$. 3D climate models predict faster sublimation[108]; however, faster sublimation rates would have no effect on our conclusions. We do not include thermal buffering from this icy material, but this would only slightly delay/damp the thermal wave. Ice-overburden sublimation tends to enhance and extend the atmosphere $CH_4$-enrichment episode in our model. To show that ice-overburden sublimation is not required for a $CH_4$ burst, the results of a high-$\varphi$ to low-$\varphi$ obliquity transition are shown in Fig. S5. In this simulation ice unloading does not occur (because nontropical ice is always unstable), and methane bursts still result.

$CH_4$-clathrate stability zone boundaries are taken from Table 4.1 of ref. 18. Destabilized $CH_4$ is assumed to be released to the atmosphere during the same timestep. Rapid release is a reasonable approximation because the thermal pulses are at orbital frequencies ($10^5$-$10^6$ yr), and – especially when fracturing associated with clathrate destabilization is taken into account – $CH_4$ is unlikely to be trapped for this long. Once released, $CH_4$ is not recharged.

We evaluated additional $CH_4$ release from taliks beneath lakes. Sub-talik $CH_4$ release is initialized once atmospheric $CH_4$ exceeds an arbitrary, but radiatively reasonable threshold of $10^2$ kg $m^{-2}$. We take this threshold to mark the onset of lake flooding. The warming-front depth is set to $2.32\sqrt{(\kappa\tau)}$ where $\kappa = 10^{-6}$ $m^2$ $s^{-1}$ is thermal diffusivity and $\tau$ is time since sub-talik release is initialized. $f$ beneath lakes is the same as $f$ elsewhere. All $CH_4$ above the warming-front is released to the atmosphere. The warming-front's progress is halted at a depth 350 m, corresponding to pressure-stabilization of $CH_4$-clathrate at ~273K. For a talik area of $1.1\times10^6$ $km^2$ (corresponding to the Eridania paleolake; ref. 35), the talik feedback is minor compared to feedback release of $CH_4$ from un-inundated locations. However, if the flooded area was larger[36,109], talik feedback could be important.

**$CH_4$ Destruction Parameterization.**
We used the Caltech/JPL 1-Dimensional Mars photochemistry code, modified to include reduced C species[23,110-111]. Estimated 2.7 Ga photon fluxes are used[26]; our results would remain qualitatively the same for photon fluxes corresponding to times <3.8 Ga. Boundary conditions include surface burial of $O_2$, $O_3$, $H_2O_2$, and CO. $H_2O$ is set to a specified mixing ratio at the surface, is well mixed up to the saturation altitude, follows the saturation vapor pressure until the atmosphere (by assumption) becomes isothermal, and becomes well mixed again above that. Results are insensitive to the specified $H_2O$ surface-mixing ratio. The model is initialized with a specified amount of $CH_4$. This initial $CH_4$, and the (fixed) $CO_2$ abundance, were both varied. Results are shown in Figs. S3-S4. Atmospheric $H_2$ levels rise as $CH_4$ is destroyed, but the radiative effect of this $H_2$ is minor compared to $CH_4$ assuming the $H_2$-$CH_4$-$CO_2$ CIA parameterization of ref. 22.

At the $CH_4/CO_2$ ratios that are most relevant for this study (~0.005-0.02), $CH_4$ destruction rate is a function of $CO_2/CH_4$ ratio. This can be simply interpreted as the result of competition between $CO_2$ and $CH_4$ for UV photons: because $CH_4$ photolysis cross section is ~$10^2\times$ that of $CO_2$ near Lyman-α wavelengths (~121.6 nm), $CO_2$ increasingly shields $CH_4$ from destruction as $CH_4/CO_2$ ratio decreases. At $CH_4/CO_2$ ratios that are less relevant for our climate scenario, more complicated behavior emerges. As expected[112], $CH_4/CO_2$ >~ 0.1 leads to significant quantities of higher hydrocarbons and these could form an antigreenhouse haze. Because the model does not track production of hydrocarbons with mass greater than $C_2H_6$, we do not attempt to track haze formation. For our $CH_4/CO_2$ = 0.1 runs, autocatalysis[113-114] can play a significant role in $CH_4$ loss. For $CH_4/CO_2$ <0.005, path dependence can be important, in that the secondary products of a high $CH_4/CO_2$ pulse interact with the H and H-species (like OH) produced by destruction of the remainder of the $CH_4$[XXX]. This appears to be the cause of the scatter at low values of $CH_4/CO_2$ in Figure S4.

The results are insensitive to varying the water volume-mixing ratio from 1 ppm to saturation at the surface. Results are sensitive to varying stratospheric diffusivity $K_{zz}$ because $K_{zz}$ regulates supply of $CH_4$ from the shielded lower atmosphere to the region of UV photolysis. For {$pCO_2$ = 500 mbar, $pCH_4$ = 1 mbar}, increasing $K_{zz}$ by a factor of 100 reduces $CH_4$ lifetime by a factor of 8. This remains consistent with geologic constraints, and 100× the nominal $K_{zz}$ profile is already pushing the limits on estimates of $K_{zz}$ in the Martian atmosphere. To test sensitivity to photon flux, we first set $K_{zz}$ to 100× nominal. Then, varying Ly-α, we found that the time-to-halving of initial $CH_4$ concentration scaled approximately as $(\text{flux})^{-0.8}$. The photon-flux dependence is likely itself dependent on $CH_4$ concentration. Ly-α flux as a function of star age can be estimated using

$$I_{1216} = (3.7\times10^{-11}\ \text{cm}^{-2}\ \text{s}^{-1}) \times t_{Gyr}^{-0.72} / 4.56^{-0.72} \tag{4}$$

where $t_{Gyr}$ is time after Mars formation in Gyr[116]. Plausible variations in this scaling could affect UV flux by a factor of several[117].

The ancient destruction of atmospheric $CH_4$ at high $pCH_4$ (Figs. S3-S4) proceeds differently to destruction of $CH_4$ at low $pCH_4$ (e.g., modern Mars). At low $pCH_4$, roughly half of $CH_4$ loss is accounted for[25] by reactions between methane and oxidizing agents from photolysis of $H_2O$ and $CO_2$. These reactions become less important as $pCH_4$ increases.

We did not find plausible parameter combinations for which $CH_4$ lifetime is <10 Kyr. Therefore, it is reasonable to infer on the basis of these results that the validity of the methane burst scenario depends on the size of the methane burst supplied to the atmosphere. If a burst is large enough to alter lake hydrology, then $CH_4$ destruction will be slow enough to match the lake-lifetime constraints.

**Data availability.**
The materials that support the findings of this study and the figures in this paper, including computer code, are available from the corresponding author upon request.

**Methods-only references.**

**SI Guide**

**Supplementary Figures and Supplementary Table**
pdf format, 2.8 MB.

The Supplementary Figures show:

(1) An olivine outcrop in an alluvial fan source region
(2) Clathrate charge-up and release scenario
(3) Methane drawdown curves varying initial $pCH_4$ and $pCO_2$
(4) Methane destruction rates plotted against $CH_4/CO_2$ ratio
(5) Detailed results for different $CH_4$-release scenarios.

The Supplementary Table shows minimum lake lifetimes obtained for 6 Mars deltas.

**Supplementary Information.**

| Site | Delta volume ($V_d$, km$^3$) | Lake area ($A_p$, km$^2$) | Evaporation rate constraint ($E$, m/yr) | $V_W/V_S$ assumed | Minimum lake lifetime (Kyr) |
|---|---|---|---|---|---|
| Eberswalde delta (1) | 6 | >410 | <1 m/yr | $10^3$ | 15 |
| SW Melas Fan "C" (2) | 3.5 | 350 | <1 m/yr | $10^3$ | 10 |
| SW Melas Fan "F" (2) | 1.3 | 350 | <1 m/yr | $10^3$ | 4 |
| Dulce Vallis (3) | 1.5 | 3008 | <1 m/yr | $10^3$ | 0.5 |
| Farah Vallis (4) | 22.5 | 3617 | <1 m/yr | $10^3$ | 6 |
| Gale Pancake (3) | 14 | 5832 | <1 m/yr | $10^3$ | 3 |

*Sources of measurements:* 1. Irwin et al. 2015. 2. Williams & Weitz 2014. 3. Palucis et al. 2016.

**Table S1.** Minimum paleolake lifetimes. We used published delta volume and lake area data, and applied a uniform lake evaporation rate and sediment:water ratio.

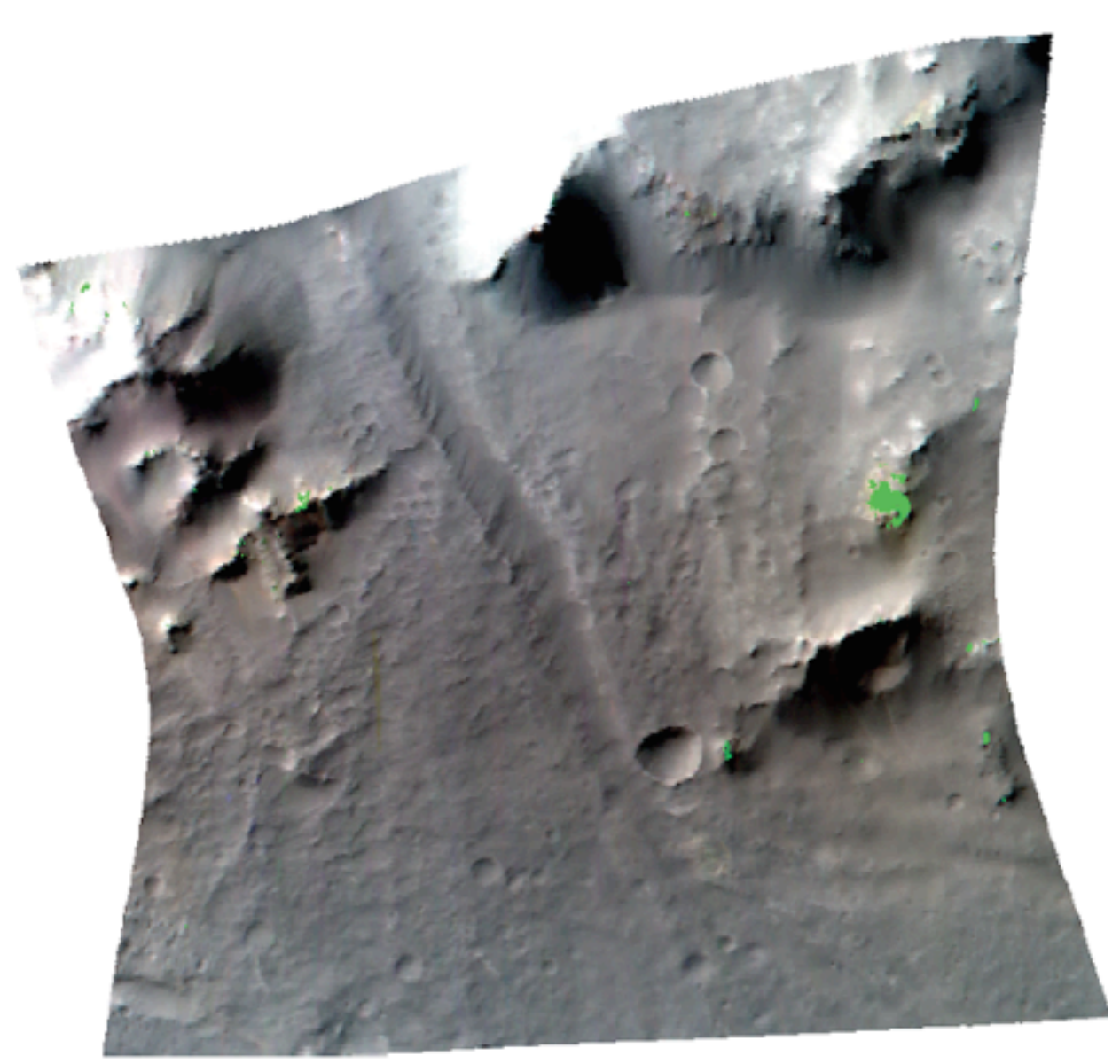

**Fig. S1.** An olivine outcrop in an alluvial fan source region (fan drains to bottom of image). Olivine detections highlighted in green. Spectra for individual pixels within these areas were checked manually in order to verify that absorptions diagnostic of olivine were present. CRISM FRT00016E79, Saheki crater. Work by David P. Mayer.

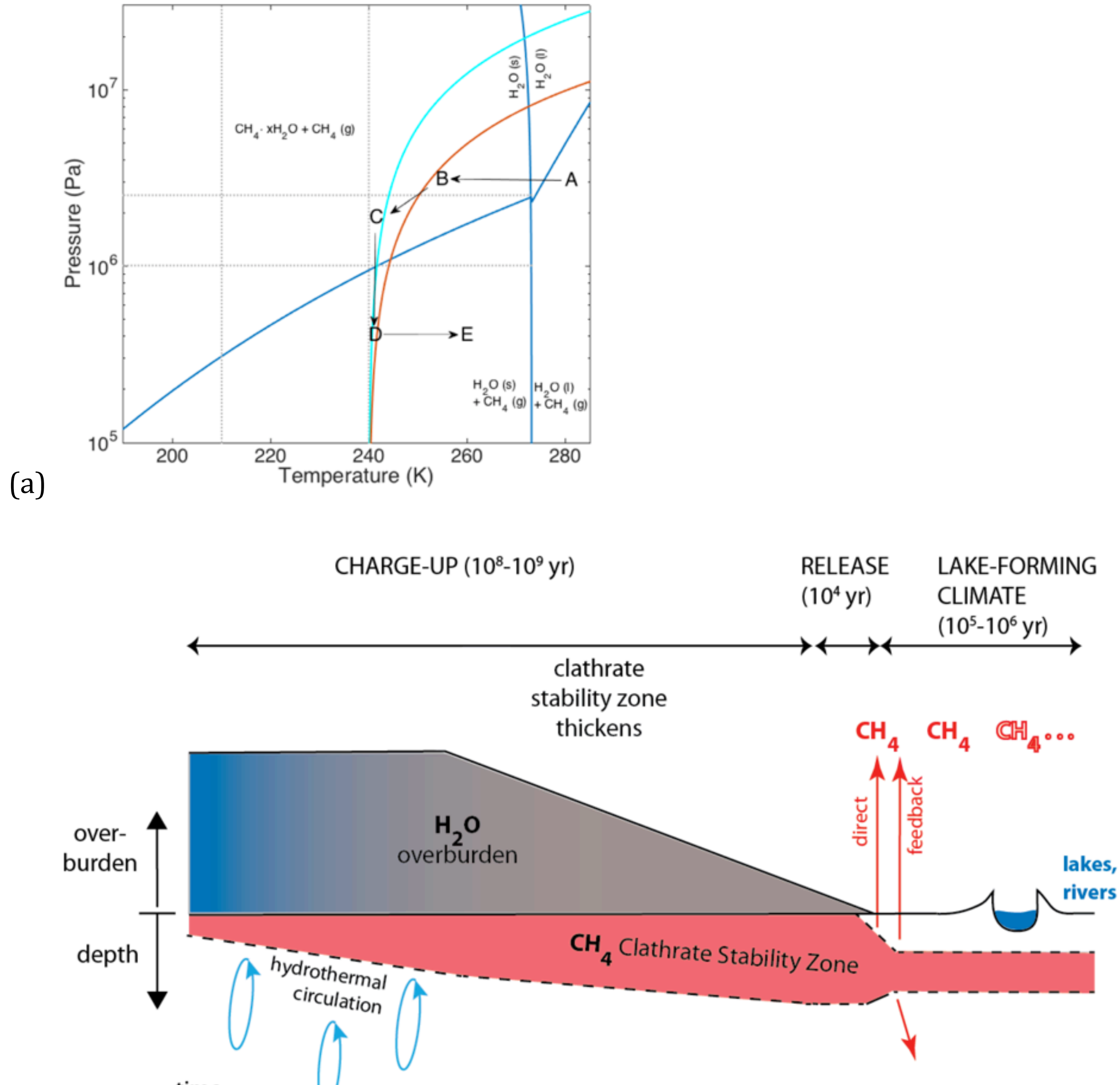

**Fig. S2.** Clathrate charge-up and release scenario. **(a)** Methane clathrate phase diagram, showing pathways to charge-up and release. Phase boundaries shown in dark blue. Mars geotherms shown in red (early, steep geotherm) and cyan (later, shallow geotherm). Early in Mars history, cooling of the geotherm locks-in $CH_4$ as clathrate in regolith, e.g. beneath early seas or ice sheets (A→B). Further geotherm cooling and escape of ice-sheet water to space (B→C) has little effect on $CH_4$-clathrate stability. Orbital change drives ice shift which leads to $CH_4$ breakdown (C→D). Orbitally-induced warming of the surface, plus warming induced by earlier release of $CH_4$ , move the regolith deeper into the $CH_4$-clathrate destabilization region (D→E). (In practice, steps C→D and D→E overlap). **(b)** Schematic of the long-term evolution of a column of the Mars uppermost crust.

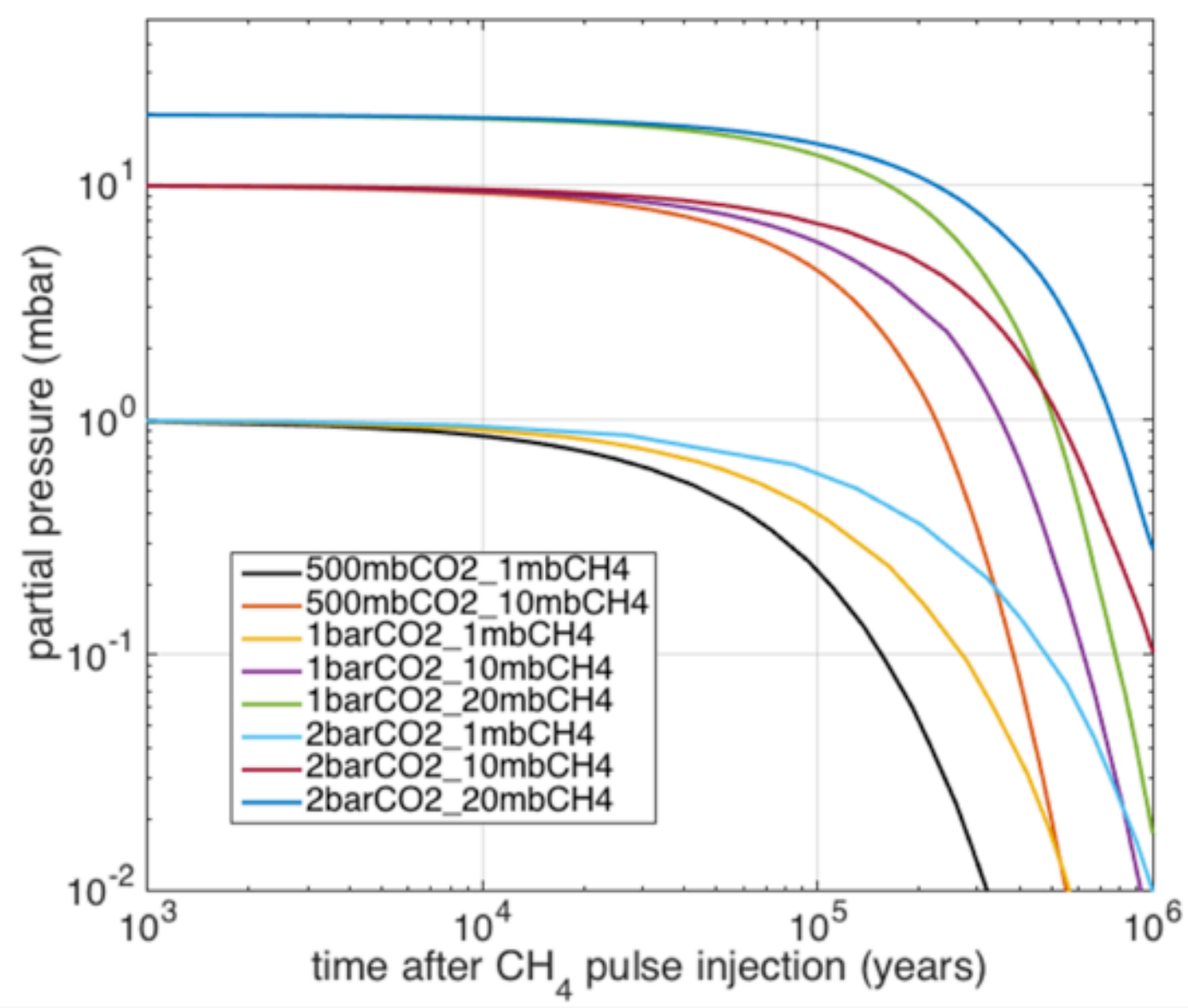

**Fig. S3.** Methane drawdown for initial methane concentrations of 1mb, 10mb, and 20mb, in $CO_2$ atmospheres of varying thickness. The 500 mb $CO_2$, 20 mb $CH_4$ case is not shown due to numerical instability.

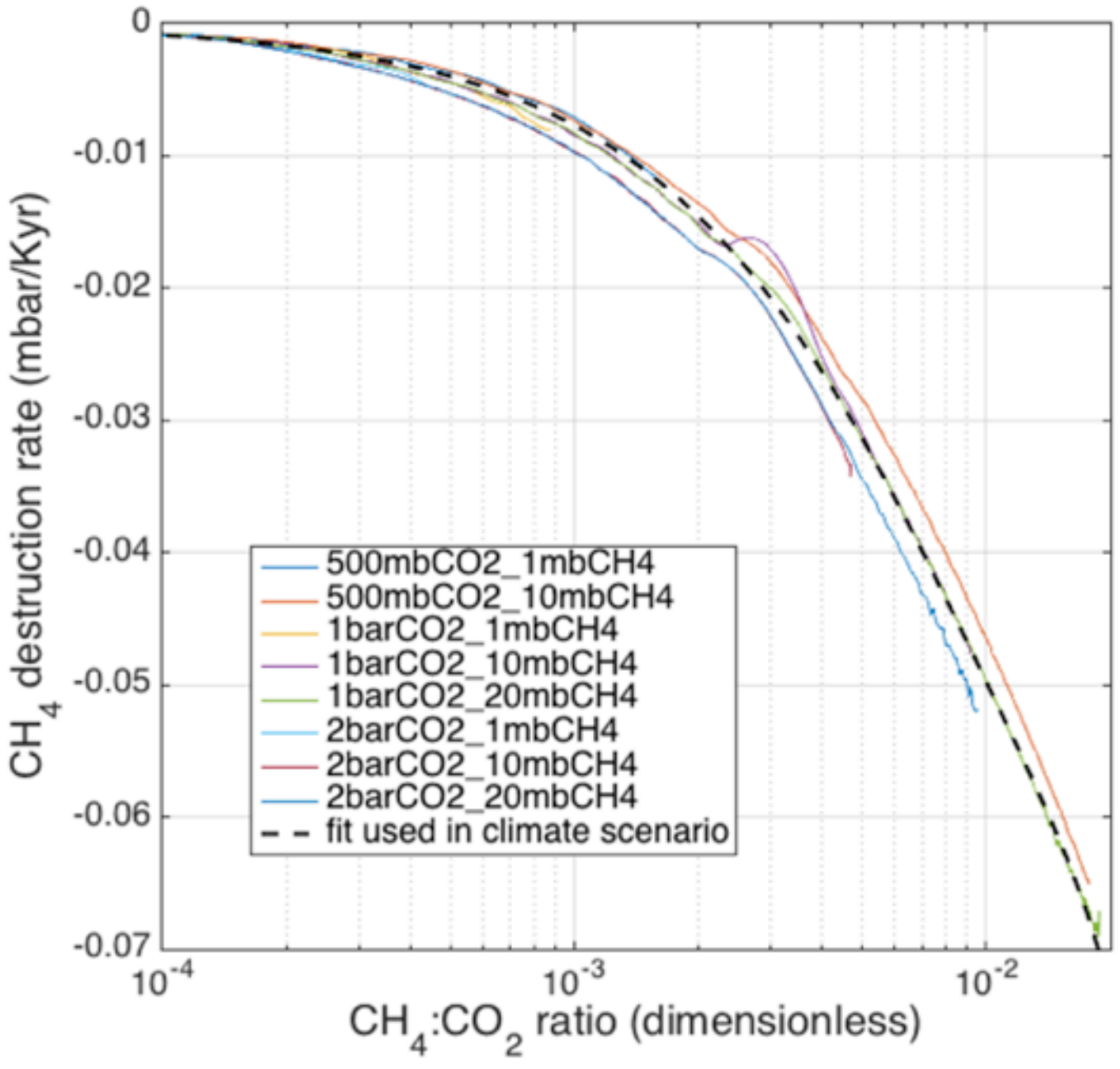

**Fig. S4.** Methane destruction rate. The first 15 Kyr of each run are excluded due to numerical artifacts associated with model startup. The dashed black line is a fit to the 1 bar $CO_2$, 20 mbar $CH_4$ run.

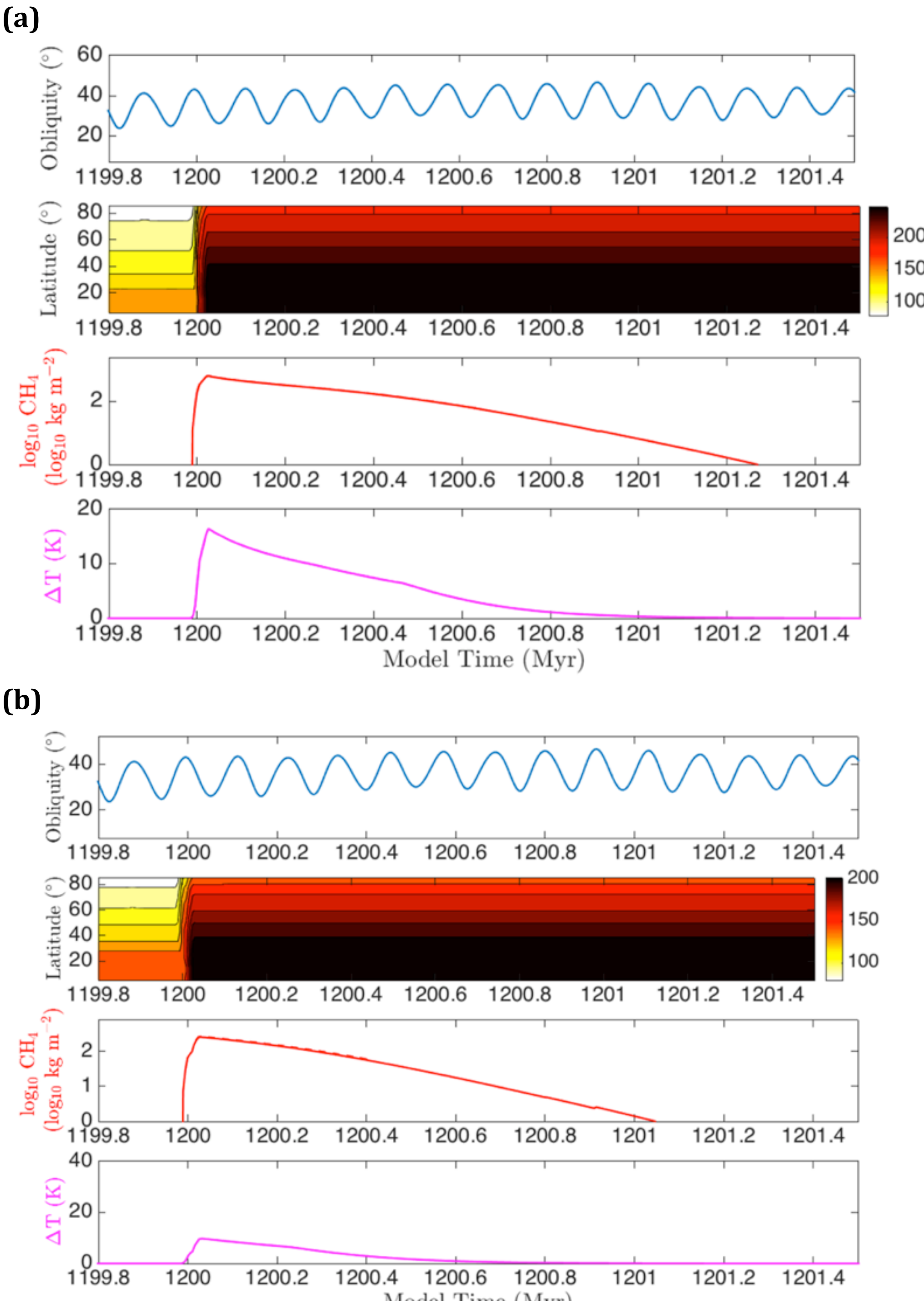

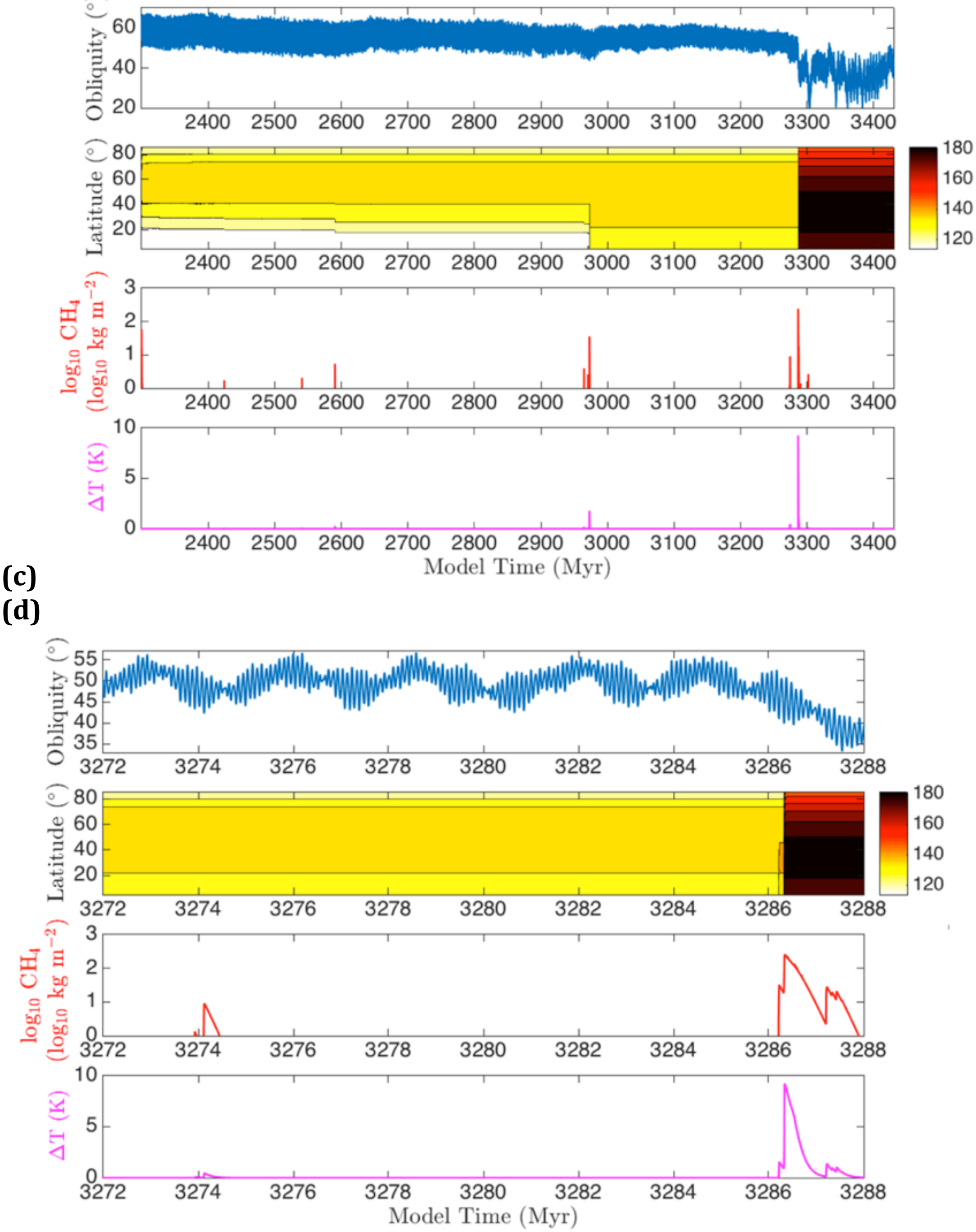

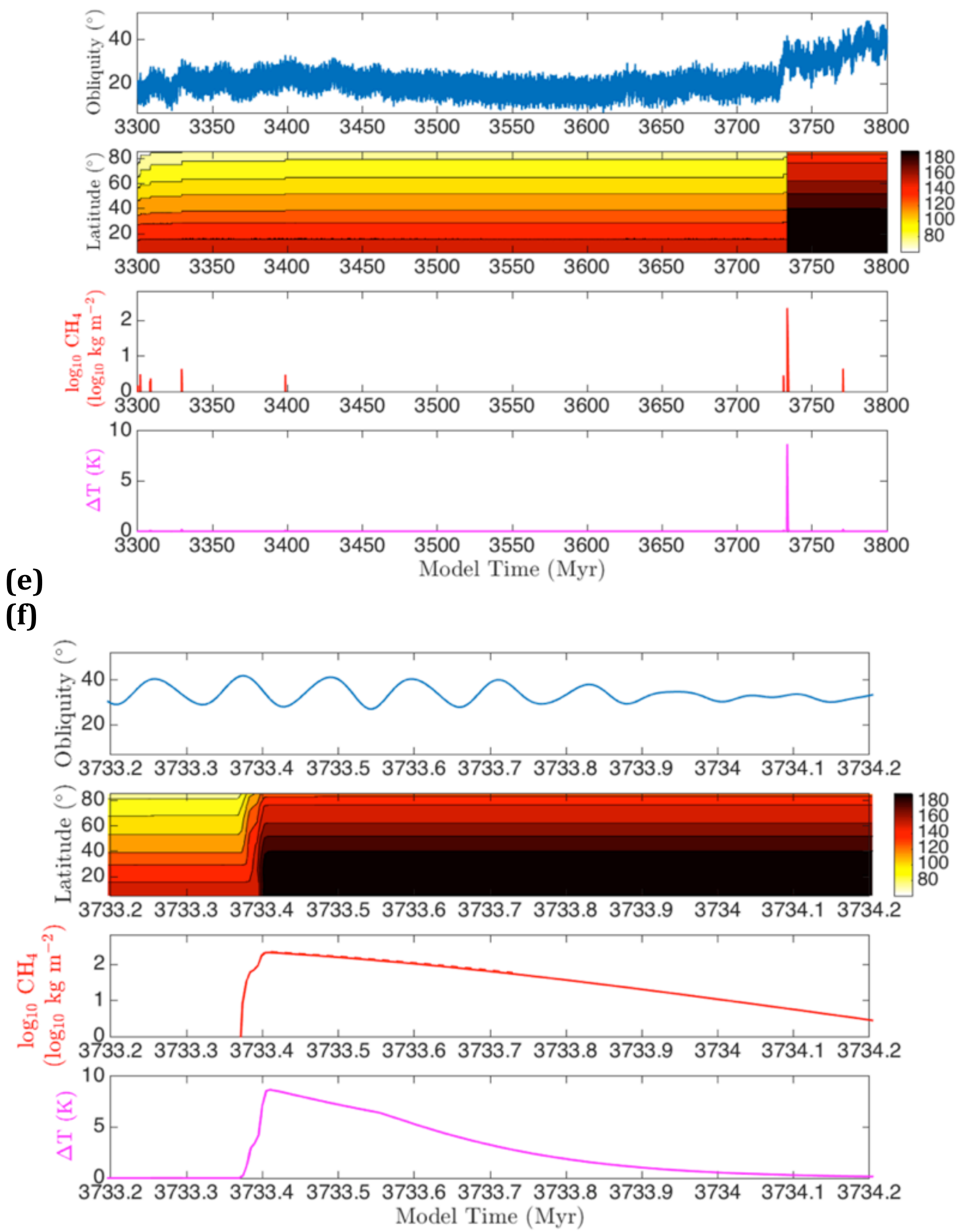

**Fig. S5.** Different $CH_4$-release scenarios. Model time is arbitrary. For each subfigure, the top panel shows example obliquity forcing. The colors in the second panel show the depth to the top of the clathrate-hydrate stability zone (depth in meters). Darkening of colors indicates clathrate destabilization. The third panel shows atmospheric $CH_4$ column mass. Dashed line includes talik feedback. The bottom panel shows temperature change. Solid line is for $CH_4$ alone; dashed line is for $CH_4$ +10%$C_2H_6$. (a) Zoom in on the biggest $CH_4$ burst from the $f$ = 0.045 simulation shown in Fig. 4. (b) As for (a), but with $f$ = 0.03, showing strong sensitivity to $f$. (c) $CH_4$ bursts for a simulation of long-term $\varphi$ decline (temperature effects only, no decompression); $f$ = 0.045. (d) Zoom in on part of (c). (e) Showing a different $\varphi$-rise scenario, with $f$ = 0.03 (compare to Fig. 4). (f) Zoom in on the biggest $CH_4$ burst from the simulation shown in (e).